\documentclass[aps,prb,onecolumn,a4paper,10pt,showpacs,citeautoscript,notitlepage,preprintnumbers]{revtex4-1}

\usepackage{graphicx,amsfonts,epsfig,latexsym,amscd,amsmath,theorem,mathrsfs,color}
\usepackage[colorlinks=true, a4paper=true, pdfstartview=FitV,linkcolor=blue, citecolor=blue, urlcolor=blue]{hyperref}

\def\comment#1{}

\newcommand{\beg}{\begin{eqnarray}}
\newcommand{\eee}{\end{eqnarray}}

\def\cm#1{}

\newcommand{\R}{{\mathbb{R}}}

\newcommand{\hh}{{\cal H}}
\newcommand{\eps}{\epsilon}

\newcommand{\f}{\frac}

\newcommand{\be}{\begin{equation}}
\newcommand{\ee}{\end{equation}}
\newcommand{\ba}{\begin{eqnarray}}
\newcommand{\ea}{\end{eqnarray}}
\newcommand{\beq}{\begin{equation}}
\newcommand{\eeq}{\end{equation}}
\newcommand{\bea}{\begin{eqnarray}}
\newcommand{\eea}{\end{eqnarray}}
\newcommand{\bastar}{\begin{eqnarray*}}
\newcommand{\eastar}{\end{eqnarray*}}

\newcommand{\cd}{\partial}

\newcommand{\ignore}[1]{}
\newcommand{\mbf}{\mathbf}
\newcommand{\Eqref}[1]{\eqref{#1}}
\newcommand{\Figref}[1]{Fig.~\ref{#1}}

\newcommand{\ie}{{\it i.e.~}}
\newcommand{\resp}{\emph{resp.~}}

\begin{document}
\preprint{Physica C : Superconductivity {\bf XX}, XXXXXX (2012)}
\title{\texorpdfstring{Type-1.5 superconductivity in multiband systems: \\  magnetic response, broken
symmetries and microscopic theory. A brief overview.}{Type-1.5 superconductivity in multiband systems:  
magnetic response, broken symmetries and microscopic theory. A brief overview.}}

\author{
E. Babaev${}^{1,2}$, J. Carlstr\"om${}^{1,2}$,  J. Garaud${}^2$, M. Silaev${}^{1,3}$ and  J.M. Speight${}^4$}
\affiliation{
${}^1$Department of Theoretical Physics, The Royal Institute of Technology, Stockholm, SE-10691 Sweden\\
${}^2$ Department of Physics, University of Massachusetts Amherst, MA 01003 USA\\
${}^3$ Institute for Physics of Microstructures RAS, 603950 Nizhny Novgorod, Russia.\\
${}^4$ School of Mathematics, University of Leeds, Leeds LS2 9JT, UK
}

\begin{abstract}

A conventional superconductor
is described by a single complex order parameter field which has
 two fundamental length scales,
the magnetic field penetration depth $\lambda$ and the coherence length $\xi$.
Their ratio  $\kappa$   determines the response
of a superconductor to an external field, sorting  them into two categories as follows; type-I when $\kappa <1/\sqrt{2}$ and type-II when $\kappa >1/\sqrt{2}$.
We overview here multicomponent systems which
can possess three or more fundamental length scales and  allow a
separate ``type-1.5" superconducting state when, e.g. in two-component
case  $\xi_1<\sqrt{2}\lambda<\xi_2$. In that state, as a
consequence of the extra fundamental length scale, vortices attract one another
at long range but repel at shorter ranges.
As a consequence the system should form an additional Semi-Meissner
state which properties we discuss below. In that state vortices form
clusters in low magnetic fields. Inside the cluster
one of the component is depleted and the superconductor-to-normal
interface has negative energy.
In contrast the current in second component
is mostly concentrated on the cluster's boundary, making
the energy of this interface positive. Here we briefly overview recent developments
in Ginzburg-Landau and microscopic descriptions of this state. {\sl Prepared for the proceedings of Vortex VII   conference, Rhodos September 2011}.
\end{abstract}
\maketitle
\section{Introduction}

Type-I superconductors expel weak magnetic fields, while strong fields give rise to formation of
 macroscopic normal domains with magnetic flux \cite{GL,deGennes,prozorov}. The response of type-II superconductors is
   different \cite{abrikosov}; below some critical value $H_{c1}$, the field is expelled. Above this value a superconductor forms a lattice
or a liquid of vortices which carry magnetic flux through the system. Only at a higher second critical value, $H_{c2}$ superconductivity is destroyed.

These different responses are usually viewed as  consequences of  the vortex interaction in these systems,
 the energy cost of a boundary between superconducting and normal states and  the thermodynamic stability
of vortex excitations.
In a type-II superconductor the
energy cost of a boundary between the normal and the superconducting state is
negative, while  the interaction between vortices is repulsive \cite{abrikosov}.
This leads to a formation of stable vortex lattices and liquids.
In type-I superconductors the situation is the opposite; the vortex interaction is attractive (thus making them
unstable against collapse into one large ``giant" vortex),
while the boundary energy between normal and superconducting
states is positive.

One can distinguish also a special ``zero measure'' boundary case where $\kappa$
has a critical value exactly at the type-I/type-II boundary,
which in the most common GL model parameterization corresponds to
 $\kappa = 1/\sqrt{2}$. In that
case vortices do not interact \cite{kramer,bogomolny} in the Ginzburg-Landau theory.
The noninteracting regime, which is frequently called ``Bogomolny limit''
is a property of Ginzburg-Landau model where, at $\kappa = 1/\sqrt{2}$,
the core-core attractive interaction between vortices exactly cancels
the current-current repulsive interaction \cite{kramer,bogomolny}.
However indeed in a realistic system even
in the limit
 $\kappa = 1/\sqrt{2}$, there will be always
  leftover inter-vortex interactions,
appearing beyond the GL field theoretic description,
 form underlying
microscopic physics.
The form of that interaction potential
is  determined  not by the fundamental length scales of the GL theory but
by  non-universal microscopic physics
and it can indeed be non-monotonic \cite{Jacobs-VortexAttraction}.
 These microscopic corrections are extremely small, however they can be relevant not only at 
$\kappa = 1/\sqrt{2}$ but also in a very narrow window of parameters
near $\kappa \approx 1/\sqrt{2}$ where
intervortex forces in GL theory are also very small.
We do not consider the  physics
which arises in the Bogomolny limits in this paper, rather concentrating
on the physics associated with fundamental modes of GL field theory.

Recently there has been
 increased interest in superconductors with several superconducting components.
 The Ginzburg-Landau free energy functional for multicomponent system
 has the form

\begin{equation}
F=\frac{1}{2}\sum_i (D\psi_i)(D\psi_i)^* + V(|\psi_i|)
+\frac{1}{2}(\nabla\times {\bf A})^2
\label{gl0}
\end{equation}
Here $\psi_i$ are complex superconducting components,
$D=\nabla + ie {\bf A}$, and $\psi_a=|\psi_a|e^{i\theta_a}$,
$a=1,2$,  and $V(|\psi_i|)$ stands for effective potential.
Depending on symmetry of the system there can also
be present mixed (with respect to components $\psi_i$)  gradient terms
(for a more detailed review see \cite{bcs2}).

The main situations where multiple superconducting components arise are
 (i) multiband superconductors \cite{suhl}-\cite{li} (where $\psi_i$ represent
condensates belonging to different bands), (ii)
mixtures of independently conserved condensates such as the projected
superconductivity in metallic hydrogen and hydrogen rich alloys \cite{ashcroft,Nature,Nature2,sublattice,herland},
(where $\psi_i$ represent electronic and protonic Cooper pairs
or deuteronic condensate)
or models of nuclear superconductors in neutron stars   interior  \cite{ns}
(where $\psi_i$ represent protonic and $\Sigma^-$ hyperonic condensates)
and (iii) superconductors with other than s-wave pairing symmetries.
The principal difference between the  cases (i) and (ii)
is the absence of the intercomponent Josephson coupling
in case of system like metallic hydrogen (ii) because there the
 condensates are independently conserved. Thus the symmetry
is $U(1)\times U(1)$ or higher.
In the case (i) multiple superconducting components
originate from Cooper pairing in different bands.
Because condensates in different bands are
not independently conserved
there is a rather generic presence of intercomponent Josephson coupling $\frac{\eta}{2}(\psi_1\psi_2^* +\psi_2\psi_1^*)$
in that case.

 \subsection{Type-1.5 superconductivity}

The possibility of a new type of superconductivity, distinct from the type-I and type-II
in multicomponent systems \cite{bs1,bcs,bcs2,moshchalkov,silaev,silaev2}
comes from the following considerations.
In principle the boundary problem  in the Ginzburg-Landau type of equations in the presence
of phase winding  is  {\it not}
 reducible
to  a one-dimensional problem in general.
Furthermore, as discussed in \cite{bs1,bcs,bcs2,garaud,silaev,silaev2}, in general
in two-component models  there are { three fundamental length scales:
magnetic field penetration length  $\lambda$ and two characteristic
length scales of the variations of the density fields $\xi_1,\xi_2$}
which renders the model impossible to parametrize in terms of a single
dimensionless parameter $\kappa$
and thus the type-I/type-II dichotomy is not sufficient for classification.
Rather, in a wide range of parameters, as a consequence of the existence
of three fundamental length scales, there is a separate superconducting regime
with $\xi_1/\sqrt{2}<\lambda<\xi_2/\sqrt{2}$. In that regime a situation is possible
where vortices have long-range attractive (due to ``outer cores" overlap), short-range
repulsive interaction (driven by current-current and electromagnetic interaction)
and form vortex clusters immersed in domains of  two-component Meissner state \cite{bs1,bcs}.
Recent experimental works \cite{moshchalkov,moshchalkov2}
proposed that this
state is realized in
 the two-band material MgB$_2$. In Ref.  \cite{moshchalkov}  this regime
was termed ``type-1.5'' superconductivity by Moshchalkov
et al. These works resulted in increasing interest in the subject
\cite{recent,daonew,recent2,recent2a}.
Recently type-1.5 superconductivity was  discussed in
context of quantum Hall effect \cite{qhe}.

If the vortices form clusters one cannot use the usual one-dimensional
argument concerning the energy of superconductor-to-normal state boundary
to classify the magnetic response of the system.
First of all, the energy per vortex in such a case
depends on whether a vortex is placed in a cluster or not.
Formation of a single isolated vortex might be energetically
unfavorable, while formation of vortex clusters
is favorable, because in a cluster where vortices are placed
in a minimum of the interaction potential, the energy
per flux quantum is smaller than that for an isolated vortex.

Thus, besides the energy
of a vortex in a cluster, there appears
an additional energy characteristic associated
with the boundary of a cluster.

We summarize the basic properties of type-I, type-II and type-1.5 regimes in
the table \ref{table1} \cite{bcs2}.

\begin{table*}
\begin{center}
\begin{tabular}{|p{3cm}||p{4.5cm}|p{4.5cm}|p{5cm}|}
\hline
 & {\bf  single-component Type-I } & {\bf single-component Type-II}  & {\bf multi-component Type-1.5}  \\ \hline \hline
{\bf Characteristic lengths scales} & Penetration length $\lambda$ \&   coherence length $\xi$ ($\frac{\lambda}{\xi}< \frac{1}{\sqrt{2}}$) & Penetration length $\lambda$ \& coherence length $\xi$ ($\frac{\lambda}{\xi}> \frac{1}{\sqrt{2}}$) & Two characteristic density variations length scales $\xi_1$,$\xi_2$ and penetration length $\lambda$,  the non-monotonic vortex interaction occurs in these systems in a large range of parameters when  $\xi_1<\sqrt{2}\lambda<\xi_2$
\\ \hline
 {\bf  Intervortex interaction} &   Attractive &    Repulsive   &Attractive at long range and repulsive at short range \\ \hline
{\bf  Energy of superconducting/normal state boundary} &    Positive    & Negative   & Under quite general conditions negative energy of superconductor/normal interface inside a vortex cluster but positive energy  of the vortex cluster's boundary  \\ \hline
{\bf The magnetic field required to form a vortex} &    Larger than the thermodynamical critical magnetic field  & Smaller than thermodynamical critical magnetic field & In different cases either (i) smaller than the thermodynamical critical magnetic field or (ii) larger than critical magnetic field for single vortex but smaller than critical magnetic field for a vortex cluster of a certain critical size
\\ \hline
{\bf  Phases in external magnetic field } & (i) Meissner state at low fields; (ii) Macroscopically large normal domains at larger fields.
First order phase transition between superconducting (Meissner) and normal states & (i) Meissner state at low fields, (ii) vortex lattices/liquids at larger fields.  Second order phase transitions between Meissner and vortex states and between vortex and normal states at the level of mean-field theory. & (i) Meissner state at low fields (ii) ``Semi-Meissner state":  vortex clusters coexisting with Meissner domains at intermediate fields (iii) Vortex lattices/liquids at larger fields.  Vortices form via a first order phase transition. The transition from vortex states to normal state is second order.
\\ \hline
{\bf  Energy E(N) of N-quantum axially symmetric vortex solutions} &    $\f{E(N)}{N}$  $<$ $\f{E(N-1)}{N-1}$ for all N. Vortices
collapse onto a single N-quantum mega-vortex &   $\f{E(N)}{N} >\f{E(N-1)}{N-1}$ for all N. N-quantum vortex
decays into N infinitely separated single-quantum vortices & There is a characteristic number N${}_c$ such
that $\f{E(N)}{N}$  $<$ $\f{E(N-1)}{N-1}$ for N $<$ N${}_c$, while $\f{E(N)}{N}$ $>$ $\f{E(N-1)}{N-1}$ for N
$>$ N${}_c$. N-quantum vortices decay into vortex clusters.
\\ \hline
\end{tabular}
\caption[Basic characteristics of superconductors]{Basic characteristics of bulk clean superconductors in type-I, type-II and type-1.5 regimes. Here the most common units are used in which the value of the GL parameter
which separates type-I and type-II regimes  in a single-component theory is $\kappa_c=1/\sqrt{2}$.
Magnetization curves in these regimes are shown on Fig. \ref{magnetization}}
\label{table1}
\end{center}
\end{table*}


\begin{figure}
\includegraphics[width=\linewidth]{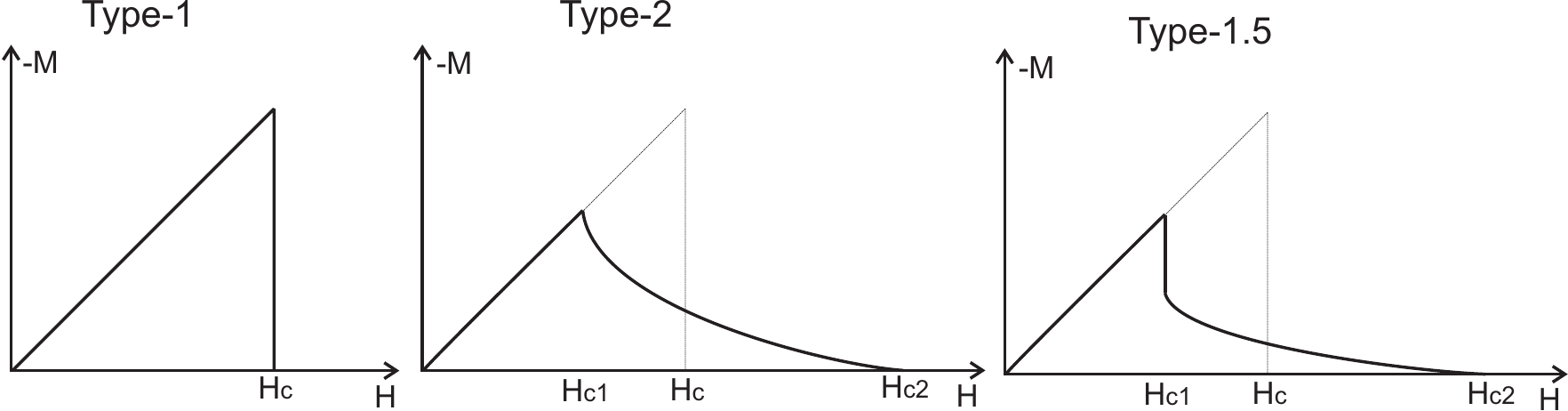}
\caption{A schematic picture of
magnetization curves of type-I, type-II and type-1.5 superconductors.
}
\label{magnetization}
\end{figure}

\subsection{Generalization to N-component case.}
The concept of type-1.5 superconductivity  has a straightforward generalization to N-component case. There it can occur in systems where characteristic length scales are $\xi_1,..,\xi_k<\sqrt{2}\lambda<\xi_{k+1},...,\xi_N$ and there are thermodynamically stable vortices with non-monotonic interaction.

\section{The two-band Ginzburg-Landau Model with   arbitrary interband
interactions. Definition of the length scales
and type-1.5 regime}
\subsection{Free energy functional}

In this section we study the type-1.5 regime using the following two-component Ginzburg-Landau (TCGL) free energy functional.

\begin{equation}
F=\frac{1}{2}(D\psi_1)(D\psi_1)^*+\frac{1}{2}(D\psi_2)(D\psi_2)^*
-\nu Re\Big\{(D\psi_1)(D\psi_2)^*\Big\}+\frac{1}{2}(\nabla\times {\bf A})^2 + F_p
\label{gl}
\end{equation}
Here
$D=\nabla + ie {\bf A}$, and $\psi_a=|\psi_a|e^{i\theta_a}$,
$a=1,2$, represent two superconducting  components  which,
 in a two-band superconductor
are associated with two different bands.
{ The term $F_p$ can contain in our analysis an {\it arbitrary} collection of non-gradient terms
representing various inter and intra-band interactions}. Below we
show how three characteristic length scales are defined in this two component model
(two associated with densities variations and
the London magnetic field penetration length).
Note that
existence of two bands in a superconductor is {\it not} a sufficient conditions for
a superconductor to be described by a model like (\ref{gl})
with two well-defined coherence lengths. Conditions of appearance of regimes
when the system does not allow
a description in terms of two-component fields
theory (\ref{gl}) is discussed in the work based on microscopic
considerations \cite{silaev,silaev2}.  
This kind of two-band GL models were also derived  earlier from microscopic two-band theories
at elevated temperature close, but not too close to $T_c$
\cite{gurevich1,zhitomirsky,gurevich2}. Although some other recent works
proposed to keep higher order gradient terms in  GL expansion \cite{shanenko},
however in Ref. \cite{silaev2} it was demonstrated that the
minimal two-band model (\ref{gl}) is microscopically justified on formal grounds and gives a very good description
of the system in a wide range of temperatures.

The only vortex solutions of the model (\ref{gl}) which have
finite energy per unit length are the integer $N$-flux quantum
vortices which have the following phase windings along a contour
$l$  around the vortex core: $\oint_l \nabla \theta_1= 2\pi N,
\oint_l \nabla \theta_2= 2\pi N$ which can be denoted as (N,N).
Vortices with differing phase windings (N,M) carry a fractional
multiple of the
 magnetic
flux quantum and have energy divergent with the system size
\cite{frac}.

In what follows  we investigate only the
integer flux vortex solutions which
are the energetically cheapest objects to produce by means of an
external field in a bulk superconductor. Note that since this object is essentially
a  bound state of two vortices, it in general will have two different co-centered cores.

\section{Intervortex forces at long range}
\label{asymp}

In this section we explain how the nature (attractive or repulsive)
 of the forces between
well separated vortices in system (\ref{gl}) can be determined purely by
analyzing $F_p$ and how three fundamental length scales can be
defined in the model (\ref{gl}).
Below we will analyze system (\ref{gl}) in the case
$\nu=0$ but for an arbitrary effective potential.  Detailed discussion of the effects of mixed gradient terms
can be found in \cite{bcs2}
By gauge invariance, $F_p$ may depend only on $|\psi_1|$, $|\psi_2|$ and
$\delta=\theta_1-\theta_2$. We consider the regime that it has a global minimum at some point
other than the one with $|\psi_a|=0$. 
We may
assume, without loss of generality, that the minimum of $F$ is at
$(|\psi_1|,|\psi_2|,\delta)=(u_1,u_2,0)$ where $u_1>0$ and $u_2\geq 0$.
Then the model has a trivial solution,
$\psi_1=u_1$, $\psi_2=u_2$, $A=0$, which we call the ground state.
It also supports vortex solutions of the form
\begin{align}
 \psi_a&=f_a(r)e^{i\theta}\,,&
(A_1,A_2)&=\frac{a(r)}{r}(-\sin\theta,\cos\theta)\label{ansatz}
\end{align}
where $f_1,f_2,a$ are real profile functions with boundary behavior
$f_a(0)=a(0)=0$, $f_a(\infty)=u_a$, $a(\infty)=-1/e$. No explicit
expressions for $f_a,a$ are known, but, by analyzing the  system
of differential equations they
satisfy, one can construct asymptotic expansions for them at large $r$,
see
\cite{bcs,bcs2}.

At large $r$ from  the vortex in the model
(\ref{gl})  the system
recovers (up to exponentially small
corrections) the ground state. In fact, the long-range field
behavior of a vortex solution can be identified
with a solution of the {\em linearization} of the model about the ground state,
in the presence of appropriate point sources at the vortex core. This idea
is explained in detail for single component GL theory in \cite{spe}.
A common feature of topological solitons (vortices being
a particular example) is that the forces they
exert on one another coincide asymptotically (at large separation) with
those between the corresponding point sources interacting via the linearized
field theory \cite{mansut}. For (\ref{gl}), the linearization has one vector
($A$) and 3 real
scalar ($\eps_1=|\psi_1|-u_1$, $\eps_2=|\psi_2|-u_2$ and $\delta$)
degrees of freedom.
The isolated vortex solutions have, by definition within the
ansatz  we use, $\delta\equiv 0$ everywhere.
Note that the GL system may also possess non-axially-symmetric solutions, such as vortex clusters,
and for these there is no reason why $\delta$ should vanish everywhere
and in fact it does not \cite{garaud}.
However below we first consider a single vortex using a axially-symmetric ansatz
, and
hence have no source for $\delta$, so we can set $\delta=0$ in the
linearization, which becomes
\be
F_{lin}=\frac12|\nabla\eps_1|^2+\frac12|\nabla\eps_2|^2+
\frac12\left(\begin{array}{c}\eps_1\\ \eps_2\end{array}\right)\cdot
\hh\left(\begin{array}{c}\eps_1\\ \eps_2\end{array}\right)
\label{gllin}
+\frac12(\cd_1A_2-\cd_2A_1)^2
+\frac12e^2(u_1^2+u_2^2)|A|^2.
\ee
Here, $\hh$ is the Hessian matrix of $F_p(|\psi_1|,|\psi_2|,0)$
about $(u_1,u_2)$, that is,
\beq
\hh_{ab}=\left.\frac{\cd^2F_p}{\cd|\psi_a|\cd|\psi_b|}\right|_{(u_1,u_2,0)}.
\eeq
Note that, in $F_{lin}$, the vector field $A$ decouples from the scalar fields
and mediates a repulsive force between vortices (originating in current-current and magnetic interaction)
with decay length which is the penetration length  $\lambda=1/\mu_A$ where $\mu_A$
is the mass of the field, that is,
\beq
\mu_A=e\sqrt{u_1^2+u_2^2}.
\eeq

By contrast, the scalar fields $\eps_1,\eps_2$ are, in general, coupled (i.e.\
in general the symmetric matrix $\hh$ has off-diagonal terms). To remove
these we  make a linear redefinition of fields, expanding
$(\eps_1,\eps_2)^T$ with respect to the orthonormal basis for $\R^2$
formed by the eigenvectors $v_1,v_2$ of $\hh$,
\beq
(\eps_1,\eps_2)^T=\chi_1 v_1+\chi_2 v_2.
\eeq
 The corresponding
eigenvalues $\mu_1^2,\mu_2^2$ are necessarily real (since $\hh$ is
symmetric) and positive (since $(u_1,u_2)$ is a minimum of $F_p$), and hence
\be
F_{lin}=\frac12\sum_{a=1}^2\left(|\nabla\chi_a|^2+\mu_a^2\chi_a^2\right)
+\frac12(\cd_1A_2-\cd_2A_1)^2
+\frac12e(u_1^2+u_2^2)|A|^2.
\ee
The scalar fields $\chi_1,\chi_2$ each mediate an attractive force between
vortices, with length scales
\begin{align}
\xi_1&\equiv 1/\mu_1 \,, &
\xi_2& \equiv 1/\mu_2 
\end{align}
 respectively.
Physically these interactions are associated with the attractive core-core
interactions. We can be somewhat more quantitative. In terms of the normal-mode fields $\chi_1,\chi_2$ and $A$, the composite point source
which must be introduced into $F_{lin}$ to produce field configurations
identical to those of  vortex asymptotics is
\begin{align}
\kappa_1&=q_1\delta(x)\,, &	
\kappa_2&=q_2\delta(x)\,, &	
{\bf j}&=m(\cd_2,-\cd_1)\delta(x)\label{cps} \,,
\end{align}
where $\kappa_1$ is the source for $\chi_1$, $\kappa_2$ the source of $\chi_2$, ${\bf j}$ the source for ${\bf A}$, $\delta(x)$ denotes the two
dimensional Dirac delta distribution and $q_1,q_2$ and $m$ are unknown real constants which can, in principle, be determined numerically by
a careful analysis of the vortex asymptotics. Physically,  a vortex, as seen
from a long distance  can be thought of as a point particle carrying two different
types of scalar monopole charge, $q_1,q_2$, inducing fields of mass $\mu_1,\mu_2$ respectively, and a magnetic dipole moment $m$ oriented orthogonal to the
$x_1x_2$ plane, inducing a massive vector field of mass $\mu_A$.
The interaction energy experienced by a pair of point particles carrying these sources, held distance $r$ apart,
is easily computed in linear field theory. For example, two scalar monopoles of charge $q$ inducing fields of mass $\mu$ held at positions
${\bf y}$ and $\tilde{\bf y}$ in $\R^2$ experience interaction energy
\be
E_{int}=-\int_{\R^2} \kappa\tilde\chi=-\int_{\R^2}q\delta({\bf x}-{\bf y})\frac{q}{2\pi}K_0(\mu |{\bf y}-\tilde{\bf y}|)
=-\frac{q^2}{2\pi}K_0(\mu|{\bf y}-\tilde{\bf y}|)
\ee
where $\kappa$ is the source for the monopole at ${\bf y}$, $\tilde\chi$ is the scalar field induced by the monopole at $\tilde{\bf y}$ \cite{spe} and
$K_0$ denotes the modified Bessel's function of the second kind. The interaction energy for a pair of magnetic dipoles may be computed similarly. In the case
of our two component GL model, the total interaction energy has three terms, corresponding to the three sources in the composite point source (\ref{cps}),
and turns out to be
\beq
E_{int}=\frac{m^2}{2\pi}K_0(\mu_A r)-\frac{q_1^2}{2\pi}K_0(\mu_1 r)-\frac{q_2^2}{2\pi}K_0(\mu_2 r).
\eeq
Note that, the first term in this formula which
originates in magnetic and current-current
interaction is repulsive, while the other two as associated
with core-core interactions of two kinds of cores are attractive. At very large $r$, $E_{int}(r)$ is dominated by whichever term
corresponds to the smallest of the three masses, $\mu_A$, $\mu_1$, $\mu_2$, so to determine whether vortices attract at long range, it is enough to compute
just these masses.

To summarize, the nature of intervortex forces at large separation can
be determined purely by analyzing $F_p$: one finds the ground state $(u_1,u_2)$
and  the Hessian $\hh$ of $F_p$
about $(u_1,u_2)$. From this one computes the mass of the
vector field $A$, $\mu_A=e\sqrt{u_1^2+u_2^2}$, and the masses
$\mu_1,\mu_2$ of the scalar normal modes
(the fields
$\chi_1,\chi_2$), these masses being the square roots of
the eigenvalues of $\hh$. If either (or both) of $\mu_1,\mu_2$ are less than
$\mu_A$, then the dominant interaction at long range is attractive (i.e.
vortex core extends beyond the area where magnetic field is localized), while
if $\mu_A$ is less than both $\mu_1$ and $\mu_2$, the dominant interaction
at long range is repulsive.
The special feature of the two-component model is that the vortices
where core extends beyond the magnetic field penetration length
are thermodynamically stable in a range of parameters and moreover one can have
a repulsive force between the vortices at shorter distances \cite{bs1,bcs,bcs2}.
It is important to stress that length scales $\mu_1^{-1},
\mu_2^{-1}$  are not directly
associated with the individual condensates $\psi_1$, $\psi_2$.
Rather they are associated with the normal modes $\chi_1,\chi_2$, defined as \cite{bcs,bcs2}
\begin{equation}
 \chi_1=(|\psi_1|-u_1)\cos\Theta-(|\psi_2|-u_2)\sin\Theta\,, ~~~
\chi_2=-(|\psi_1|-u_1)\sin\Theta-(|\psi_2|-u_2)\cos\Theta\,.
\end{equation}
These may be thought of as
rotated (in field space) versions of $\eps_1=|\psi_1|-u_1$,
$\eps_2=|\psi_2|-u_2$. The {\em mixing angle}, that is, the angle between
the $\chi$ and $\eps$ axes, is $\Theta$, where the eigenvector
$v_1$ of $\hh$ is $(\cos\Theta,\sin\Theta)^T$. This, again, can be determined
directly from $\hh$.

Note also that the shorter of the length scales  $\mu_1^{-1},
\mu_2^{-1}$, although being a fundamental length scale of the theory,
can be masked in a density profile of a vortex solution by nonlinear effects.
This, for example 
certainly happens if   $\mu_1^{-1} \ll \mu_A \equiv \lambda^{-1}$ (see short discussion in Ref. \cite{bcs2}). 
Also note that in general the minimum of the interaction potential will not be located  at the London penetration length, because it in general will be also affected by nonlinearities.

\subsection{Passive band superconductors}

To illustrate the analysis presented above, we consider the simple case of a
two band superconductor where one of the bands is passive, that is, with a
potential of the form
\beq
F_p=-\alpha_1|\psi_1|^2+\frac{\beta_1}{2}|\psi_1|^2+\alpha_2|\psi_2|^2
-\gamma(\psi_1\overline\psi_2+\overline\psi_1\psi_2)
\eeq
where $\alpha_j,\beta_1,\gamma$ are positive constants. Then $F_p$ is 
minimized when $\psi_1$ and $\psi_2$ have equal phase, and have moduli
\beq
|\psi_1|=u_1=\sqrt{\frac{\alpha_1}{\beta_1}\left(1+\frac{\gamma^2}{\alpha_1\alpha_2}\right)},\qquad
|\psi_2|=u_2=\frac{\gamma}{\alpha_2}u_1.
\eeq
The mass of the vector field $A$ is 
\beq
\mu_A=e\sqrt{u_1^2+u_2^2}=eu_1\sqrt{1+\frac{\gamma^2}{\alpha_2^2}}.
\eeq
The Hessian matrix of $F_p$ about $(u_1,u_2)$ is
\beq
\hh=\left(\begin{array}{cc}4\alpha_1+\frac{6\gamma^2}{\alpha_2}&-2\gamma\\
-2\gamma&2\alpha_2\end{array}\right).
\eeq
It is straightforward to compute explicit expressions for the eigenvalues
$\mu_1^2,\mu_2^2$ of this matrix. These are somewhat complicated, but
power series expansion in $\gamma$ reveals that
\beq
\mu_1=2\sqrt{\alpha_1}+O(\gamma^2),\qquad
\mu_2=\sqrt{2\alpha_2}+O(\gamma^2).
\eeq
Similarly, the normalized eigenvector associated with eigenvalue $\mu_1^2$ is
\beq
v_1=\left(\begin{array}{c}1\\-(2\alpha_1-\alpha_2)^{-1}\gamma\end{array}\right)
+O(\gamma^2)\eeq
so the normal modes of fluctuation about the ground state are rotated through
a mixing angle 
\beq
\Theta=-(2\alpha_1-\alpha_2)^{-1}\gamma+O(\gamma^2).
\eeq
In comparison with the uncoupled model ($\gamma=0$) then, we see that, for
small coupling $\gamma$ the length scales $1/\mu_A,1/\mu_1,1/\mu_2$ are
unchanged to leading order, but the normal modes with which $1/\mu_1,1/\mu_2$
are associated are mixed to leading order. In particular, there are
large regions of parameter space where $\mu_2<\mu_A<\mu_1$, so that
vortices attract at long range, even though the active band, $\psi_1$, is
naively ``type~II'' (that is, $\beta_1>e^2/4$).

\section{Vortex clusters in a Semi-Meissner state and non-pairwise intervortex forces.}

\subsection{Model}

In this section, following Ref. \cite{garaud} we consider in more detail two-component Ginzburg-Landau models, with
and without Josephson
 coupling $\eta$ which  directly couples the two condensates  (for treatment of other kinds of
interband couplings see \cite{bcs2}). When $\eta=0$ the condensates are
coupled electromagnetically.
\begin{align}
\mathcal{F}&=
\frac{1}{2}\sum_{i=1,2}\Biggl[|(\nabla+ ie{\bf A}) \psi_i  |^2
+ (2\alpha_i+\beta_i|\psi_i|^2)|\psi_i|^2\Biggr]
+\frac{1}{2}(\nabla \times {\bf A})^2 -\eta|\psi_1|| \psi_2|\cos(\theta_2-\theta_1)
\label{FreeEnergy}
\end{align}
Here again, the gauge covariant derivative is  $D=\nabla+ie{\bf A}$, and $\psi_i=|\psi_i| e^{i\varphi_i}$
are complex fields representing the superconducting components.
As will be discussed in Sec.\ref{microscopic},
calculations based on microscopic two-band Eilenberger theory show that the model
\Eqref{FreeEnergy} can be used to study vortex physics in multiband superconductor for a wide range
or parameters.

We also discuss importance of  complicated non-pairwise forces between superconducting vortices arising
in certain cases in multicomponent systems \cite{garaud,remark9}. These non-pairwise forces in certain cases have important consequences for vortex clusters
formation in the type-1.5
regime.

\subsection{Vortex clusters in a Semi-Meissner state and non-pairwise interactions.}

In this section we allow fluctuations in phase difference. When there is non-zero interband Josephson coupling. The phase difference is associated with a  massive mode. Its mass is
 $\sqrt{\eta (u_1^2+ u_2^2) /u_1 u_2}$.

\begin{figure}[!htb]
  \hbox to \linewidth{ \hss
 \includegraphics[width=\linewidth]{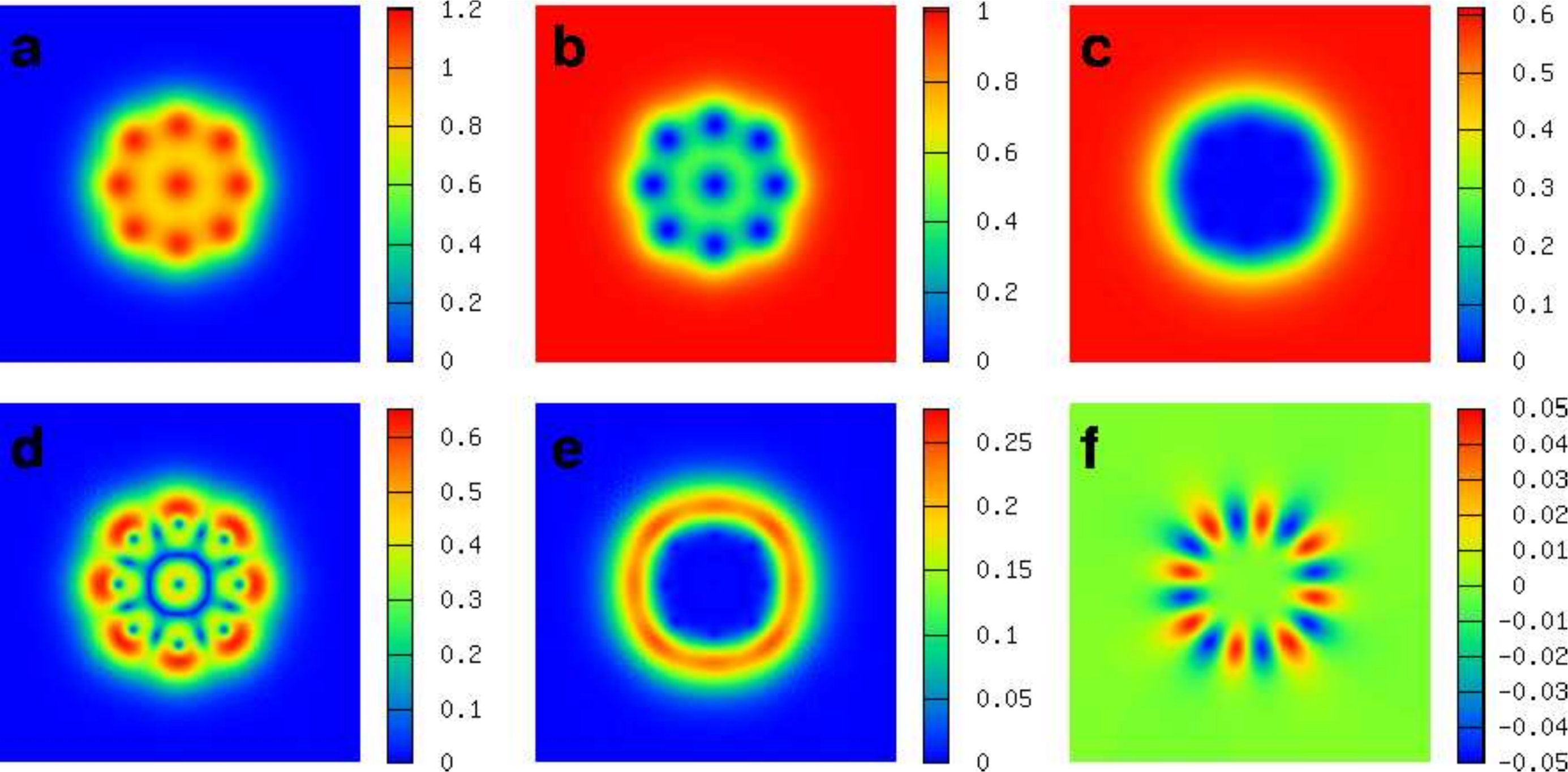}
 \hss}
\caption{
Ground state of $N_v=9$ flux quanta in a type-1.5 superconductor enjoying $U(1)\times U(1)$
symmetry of the potential (\ie $\eta=0$). The parameters of the potential being
here $(\alpha_1,\beta_1)=(-1.00, 1.00)$ and $(\alpha_2,\beta_2)=(-0.60, 1.00)$, while the electric
charge is $e=1.48$.
The displayed physical quantities are $\bf a$ the magnetic flux density,
$\bf b$ (\resp $\bf c$) is the density of the first (\resp second) condensate $|\psi_{1,2}|^2$.
$\bf d$ (\resp $\bf e$) shows the norm of the supercurrent in the first (\resp second) component.
Panel $\bf f$ is $\mathrm{Im}(\psi_1^*\psi_2)\equiv|\psi_1|| \psi_2|\sin(\theta_2-\theta_1)$ being nonzero
when there
appears a difference between the two condensates. Parameters are chosen so that the second component
has a type-I like behavior while the first one tends to from well separated vortices.
The density of the second band is depleted in the vortex cluster and its current is mostly concentrated
on the boundary of the cluster (see Ref.\cite{garaud}).
}
\label{2A-1}
\end{figure}

\begin{figure}[!htb]
  \hbox to \linewidth{ \hss
 \includegraphics[width=\linewidth]{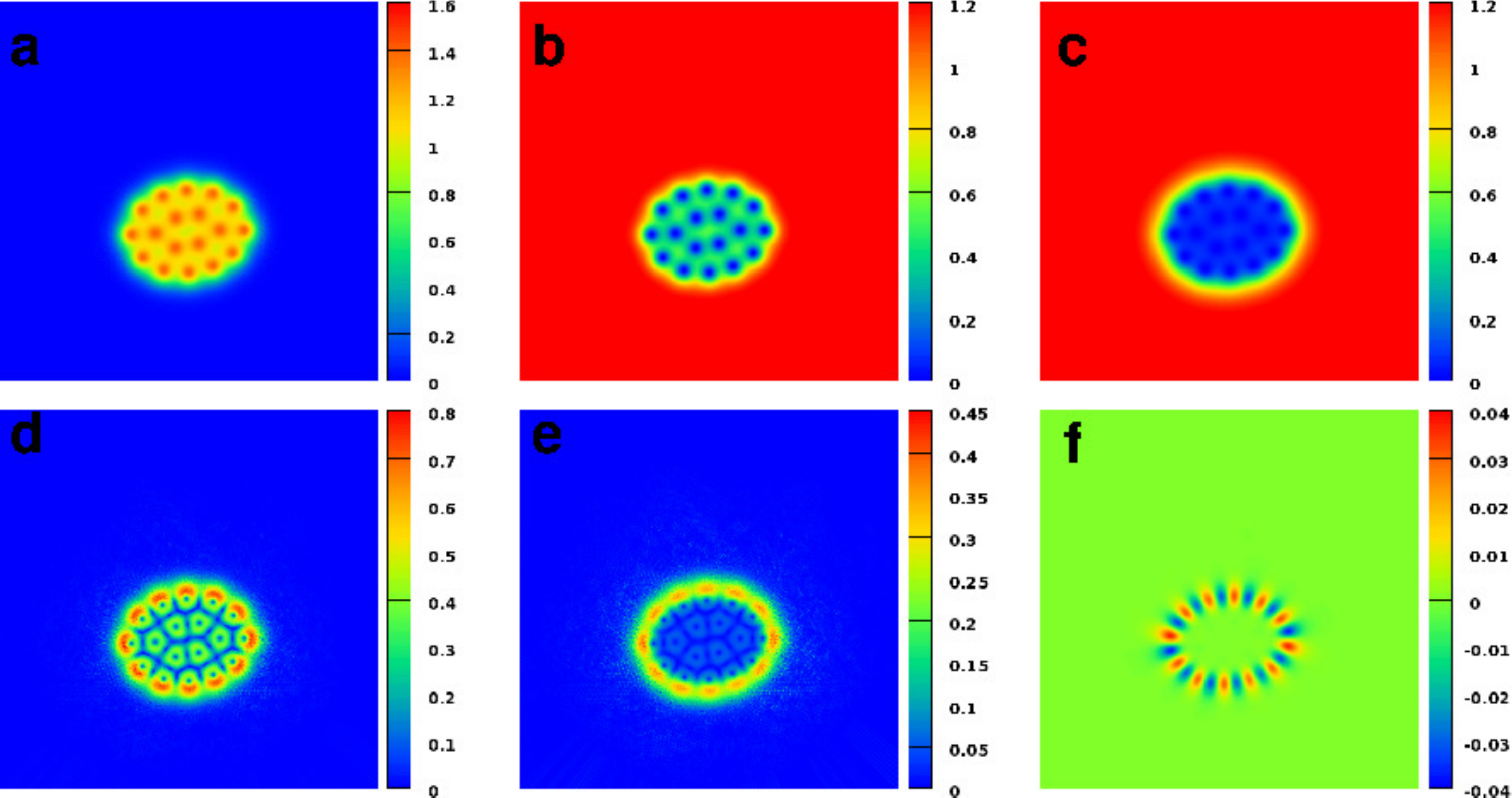}
 \hss}
\caption{
Elongated ground state cluster of $18$ vortices in a superconductor with two active bands. Parameters of the interacting potential
are $(\alpha_1,\beta_1)=(-1.00, 1.00)$, $(\alpha_2,\beta_2)=(-0.0625, 0.25)$ while the interband coupling is $\eta=0.5$.
The electric charge, parameterizing the penetration depth of the magnetic field, is $e=1.30$
so that the well in the nonmonotonic interacting potential is very small. In this case
there is visible admixture of the current of  second component in vortices inside the cluster, though
its current is predominantly concentrated on the boundary of the cluster.
}
\label{fig-new4}
\end{figure}

\Figref{2A-1} and \Figref{fig-new4} show 
numerical solutions for N-vortex bound states in several regimes (for technical details see Appendix of~\cite{garaud}).
A Nonlinear Conjugate Gradient scheme was used to solve the finite element formulation of the variational
problem, within the framework provided by the Freefem++ library~\cite{Freefem}.
Animations showing the evolution of the system, during the numerical
energy minimization, from the various  initial configurations to
the vortex clusters in the energy minimization process can be found online~\cite{julien}.
The common aspect of the regimes shown on these figures is that the density of
one of the components is depleted in the vortex cluster and
has its current most concentrated on the boundary of the vortex cluster (i.e.
has a type-I like behavior). At the same time the second component
forms a full-fledged vortex lattice inside the vortex cluster (i.e. has a type-II-like
behavior).

\subsection{Non-compact vortex clusters and non-pairwise interactions}
\begin{figure}[!htb]
  \hbox to \linewidth{ \hss
 \includegraphics[width=\linewidth]{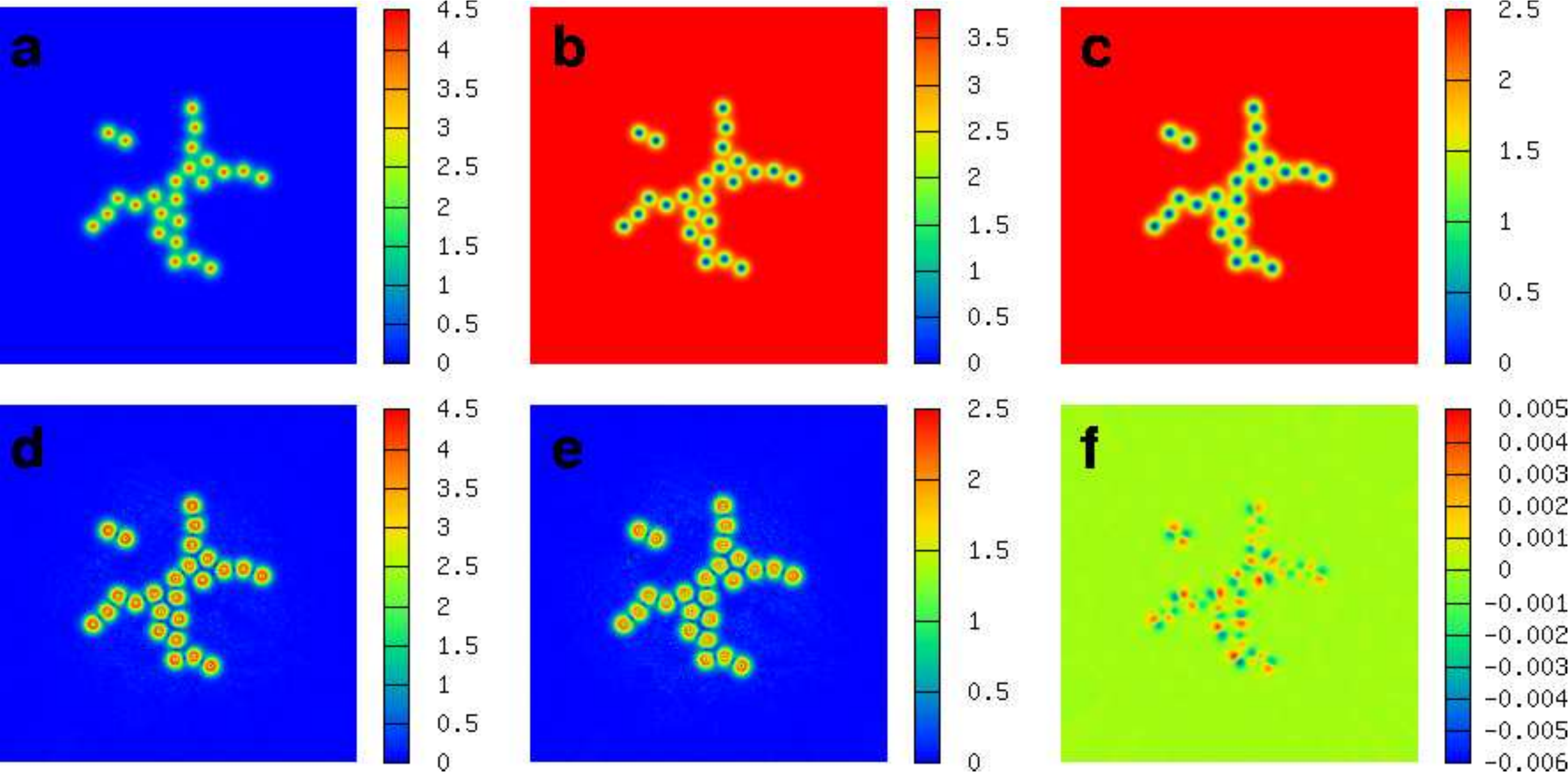}
 \hss}
\caption{A  bound state of an $N_v=25$ vortex configuration in case when
superconductivity in the second band is due to interband proximity
effect
and the  Josephson coupling is very strong $\eta=7.0$.
The initial configuration in this simulation was a giant vortex. Other
parameters
are $(\alpha_1,\beta_1)=(-1.00, 1.00)$, $(\alpha_2,\beta_2)=(3.00,
0.50)$,  $e=1.30$.
For the  simulations, like the one shown on Fig. \ref{fig-new4}  the stopping criterion
of energy minimization was when  relative variation of the norm of the gradient of the
GL functional
with respect to all degrees of freedom to be less than $10^{-6}$.
 Here
the situation is
slightly different from that shown on two previous figures. Clearly in the shown above
configuration the ground state  was not reached.
However the interaction potentials a such that
the evolution at the later stages becomes extremely slow.
The number of
energy minimization steps in this case was
order of magnitude larger than what was
required for convergence in the previous regimes.
}
\label{1A1PSJb}
 \end{figure}

Next we report the regime 
where the  passive  second band (i.e. with positive $\alpha_2$) is coupled
to the first band by extremely strong Josephson coupling $\eta=7.0$ (shown on \Figref{1A1PSJb}). This coupling
imposes  a strong energy penalty both for disparities of the condensates variations and for the difference
between phases of the condensates. Besides that in that regime there is a relatively strong non-pairwise
interactions
which diminish the energetic benefits of a triangle-like states compared to line-like vortex states \cite{garaud}.
We get a flat and complicated energy landscape and the outcome of the
energy minimization strongly depends on initial configuration.
Simulations whose outcome is compact clusters like \Figref{2A-1} and
\Figref{fig-new4} clearly ground states, since various initial guesses
lead to similar final configurations. Simulating systems like in \Figref{1A1PSJb}
is less straightforward. Numerical evolution in these systems is extremely slow
because of the very complicated energy landscape. The final state strongly depends
on the initial field configuration, indicating the configuration is not ground state but
a bound state with a very slow evolution.
Formation of highly disordered states and vortex chains due to
short-range nature of the attractive potentials and many-body forces was a common outcome of the
simulation in the similar type-1.5 regimes with strong Josephson coupling, in spite
of negligible effects of ultra-fine numerical grid.

The \Figref{1A1PSJb} shows the typical non-universal outcome of the energy minimization in
this case. Striking feature here is  formation of vortex stripe-like configuration.
Indeed it strongly contradicts the ground state expected from the two-body forces in this system.
Namely the axially symmetric two-body potentials with long range attraction and short-range
repulsion (which we have in this case) do not allow stripe formation in the ground state configurations.
Nonetheless such structures are physically entirely  possible: first the exponentially
decaying short-range forces result in very tiny forces which can drive evolution from a stripe to
compact cluster configuration. Secondly there are repulsive multibody intervortex forces
which compromise system's ability for form a compact configuration \cite{garaud}.
So in that regime vortex stripes and small lines can easily form because of non-pairwise intervortex interactions.

Note that even in this regime, the system exhibits self-induced gradients of the phase
difference, in spite of the  strong Josephson coupling.

\section{Semi-Meissner state and broken symmetries.}
As discussed above a system with non-monotonic intervortex
interaction potentials allow a state
with macroscopic phase separation
in vortex droplets and Meissner domains.
In type-1.5 superconductors this state
can also represent  a phase separation into domains
of states with different broken symmetries.
In this section we will give two different examples of
how such behavior can arise.

Note that in multicomponent superconductors some symmetries are global (i.e.
associated with the degrees of freedom decoupled from vector potential)
 and some are local  i.e. associated with the degrees of freedom
 coupled to vector potential. It is well known that in the later case the concept of spontaneous symmetry breakdown  is not defined
 the same way as in a system with global symmetry. However below, for brevity
 we will not be making terminological distinctions between local and global symmetries
(detailed discussion of these aspects can be found in  e.g. \cite{frac,Nature,herland}).

\subsection{\texorpdfstring{Semi-Meissner state as a macroscopic phase separation in $U(1)\times U(1)$ and $U(1)$ domains.}{Semi-Meissner state as a macroscopic phase separation in U(1)xU(1) and U(1) domains.}}

Consider a  superconductor with broken $U(1) \times U(1)$
symmetry, i.e.  a collection of independently conserved
condensates with no intercomponent Josephson coupling. As
discussed above, in the Semi-Meissner state in  vortex droplets
the superconducting component which has vortices with larger cores
is more depleted. In $U(1) \times U(1)$ system the vortices with
phase windings in different condensates are bound
electromagnetically which gives asymptotically logarithmic
interaction potential with prefactor proportional to
$|\psi_1|^2|\psi_2|^2/(|\psi_1|^2+|\psi_2|^2)$ \cite{frac}, and
even weaker interaction strength at shorter separations.

Consider now a macroscopically  large vortex domain. Even if the
second component there is not completely depleted, its density is
suppressed and as a consequence the binding energy between
vortices with different phase windings ($\Delta \theta_1=2\pi,
\Delta \theta_2=0$) and  ($\Delta \theta_1=0, \Delta
\theta_2=2\pi$) can be arbitrarily small. Moreover the vortex
ordering energy in the  component with more depleted density will
also small. As a result, even small thermal fluctuation can drive
vortex sublattice melting transition \cite{Nature,sublattice}in
the macroscopically large vortex droplet.  In that case the
fractional vortices in weaker component tear themselves off the
fractional vortices in strong component and form a disordered
state. Note that the vortex sublattice melting is associated with
the  phase transition from $U(1)\times U(1)$ to $U(1)$ state
\cite{Nature,sublattice}. I.e. that vortex cluster will represent
a domain of $U(1)$ phase (associated with the superconducting
state of strong component) immersed in domain of vortexless
$U(1)\times U(1)$ Meissner state. If the magnetic field is
increased  the system will go from the Semi-Meissner state
(with coexisting $U(1)\times U(1)$ and $U(1)$ domains) to $U(1)$
vortex state.

\subsection{\texorpdfstring{Semi-Meissner state as a macroscopic phase separation in $U(1)$ and $U(1)\times Z_2$ domains
in three band type-1.5 superconductors.}{Semi-Meissner state as a macroscopic phase separation in U(1) and U(1)x Z2 domains
in three band type-1.5 superconductors.}}

In the previous subsection we considered the case where Semi-Meissner state represents
coexistences of domains with different broken symmetries as a consequence
of vortex sublattice melting  transition.
In type-1.5 systems the coexistence of domains with different broken
symmetry can also take place also in the ground state, i.e. without
the need of thermal fluctuations \cite{3bands}.
In this subsection  we discuss an example studied
in Ref. \cite{3bands}  of such behavior in a
three-band model with ``phase frustration".

The minimal GL free energy functional to model a three-band superconductor  is
\begin{equation}
 F= \frac{1}{2}(\nabla \times \mbf A)^2+ \sum_{i =1,2,3}\frac{1}{2}|\mbf D\psi_i|^2
+\alpha_i|\psi_i|^2+\frac{1}{2}\beta_i|\psi_i|^4
+\sum_{i =1,2,3}\sum_{j>i}\eta_{ij}|\psi_i||\psi_j|\cos(\varphi_{ij}) \,.
\label{freeEnergy}
\end{equation}
Here 
the phase difference between two condensates are denoted $\varphi_{ij}=\varphi_j-\varphi_i$.

 Systems with more than two Josephson-coupled bands can exhibit
 \emph{phase frustration} \cite{3bands,iron2,recent2a}. For $\eta_{ij}<0$, a given Josephson interaction energy term is minimal for zero phase difference
 (we then refer to the coupling as ``phase-locking" ), while when  $\eta_{ij}>0$ it is minimal for a phase difference equal to $\pi$
 (we then refer to the coupling as ``phase-antilocking" ). Two component systems are symmetric with respect to the sign change
$\eta_{ij}\to -\eta_{ij}$ as the phase difference changes by a factor $\pi$, for the system to recover the same interaction. However, in
systems with more than two bands there is generally no such symmetry. For example if a three band system has $\eta>0$ for all
Josephson interactions, then these terms can not be simultaneously minimized, as this would correspond to all
possible phase differences being equal to $\pi$.

The ground state values of the fields  $|\psi_i|$ and $\varphi_{ij}$ of system \Eqref{freeEnergy} are found by minimizing its potential
energy
\be
\sum_i\Big\{\alpha_i|\psi_i|^2+\frac{1}{2}\beta_i|\psi_i|^4\Big\}
+\sum_{j>i}\eta_{ij}|\psi_i||\psi_j|\cos(\varphi_{ij}).
\label{potential}
\ee
Minimizing the potential energy \Eqref{potential} can not in general be done analytically. Yet, some properties
can be derived from qualitative arguments. In terms of the sign of the $\eta$'s, there are four principal situations:

\begin{center}
\begin{tabular}{ c||c|cc }
Case & Sign of $\eta_{12},\eta_{13},\eta_{23}$ & Ground State Phases \\
\hline
1& $- - -$ & $\varphi_1=\varphi_2=\varphi_3$ \\
2& $- - +$ & Frustrated  \\
3& $- + +$ & $\varphi_1=\varphi_2=\varphi_3+\pi$ \\
4& $+ + +$ & Frustrated
 \end{tabular}
\end{center}

The case 2) can result in several ground states. If $|\eta_{23}|\ll |\eta_{12}|,\;|\eta_{13}|$, then the phase differences
are generally $\varphi_{ij}=0$. If on the other hand $|\eta_{12}|,\;|\eta_{13}| \ll |\eta_{23}| $ then $\varphi_{23}=\pi$
and $\varphi_{12}$ is either $0$ or $\pi$. For certain parameter values it can also have compromise states with
$\varphi_{ij}$ not being integer multiples of $\pi$.

\begin{figure}[!htb]
 \hbox to \linewidth{ \hss
\includegraphics[width=0.6\linewidth]{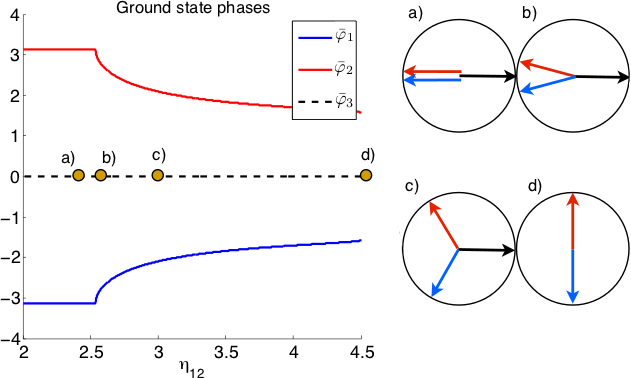}
 \hss}
\caption{
Ground state phases of the three components as function of $\eta_{12}$ (here $\varphi_3=0$ fixes the gauge).
The GL parameters are $\alpha_i=1,\;\beta_i=1,\;\eta_{13}=\eta_{23}=3$. For intermediate values of $\eta_{12}$ the ground state
exhibits discrete degeneracy (symmetry is $U(1)\times Z_2$ rather than $U(1)$) since the energy is invariant under the sign
change $\varphi_2\to-\varphi_2,\; \varphi_3\to-\varphi_3$. For large $\eta_{12}$ we get $\varphi_2-\varphi_3=\pi$
implying that $|\psi_3|=0$ and so there is a second transition from $U(1)\times Z_2$ to $U(1)$ and only two bands at the point d).
Here, the phases were computed in a system with only passive bands, though systems with active bands exhibit the same
qualitative properties except for the transition to $U(1)$ and two bands only (\ie active bands have non-zero density in the ground state).
}
\label{case4}
\end{figure}

The case 4) can give a wide range of ground states, as can be seen in \Figref{case4}. As $\eta_{12}$ is scaled,
ground state phases change  from  $( -\pi,\; \pi,\; 0)$ to the limit where one band is depleted and
the remaining phases are $(-\pi/2,\;\pi/2)$.

An important property of the potential energy \Eqref{potential} is that
if any of the phase differences $\varphi_{ij}$ is not an integer multiple of $\pi$,
then the ground state posses an additional discrete $Z_2$ degeneracy. For example for a system with
$\alpha_i=-1,\;\beta_i=1$ and $\eta_{ij}=1$, two possible ground state are given by
$\varphi_{12}=2\pi/3,\;\varphi_{13}=-2\pi/3$ or $\varphi_{12}=-2\pi/3,\;\varphi_{13}=2\pi/3$.
 Thus in this case, the symmetry  is $U(1)\times Z_2$, as opposed to $U(1)$.
As a result, like any other system with $Z_2$  degeneracy, the theory allows an additional set of topological excitations :
domain walls interpolating between the two inequivalent ground states.

The ground state of a phase frustrated superconductor is in many cases non-trivial, with phase differences being
compromises between the various interaction terms. Inserting vortices in such a system can shift the balance between
different competing couplings, since
vortices can in general have different effects on the different bands. In particular, since the core sizes of vortices
are not generally the same in all bands, vortex matter will typically deplete some components more than others and
thus can alter the preferred  values of the phase difference. So the minimal potential energy inside a vortex lattice or
cluster may correspond to a different set of phase differences than in the vortexless ground state.
To see this, consider the following argument:
The phase-dependent potential terms in the free energy \Eqref{freeEnergy} are of the form
\be
\eta_{ij} u_iu_j f_i(\mbf r)  f_j(\mbf r)\cos(\varphi_{ij}(\mbf r))\,,
\ee
where $u_i$ are ground state densities and each $f_i(\mbf r)$ represent an Ansatz which models how superfluid densities
are modulated due to vortices. Consider now a  system where N vortices are uniformly distributed in a domain $\Omega$.
The phase dependent part of the free energy is
\be
U_\varphi=\left[\sum_{i>j}\eta_{ij} u_iu_j \right]
  \int_{\Omega} d\mbf r  f_i(\mbf r)  f_j(\mbf r)\cos(\varphi_{ij}(\mbf r)).\label{fullrenorm}
\ee
If $\varphi_{ij}$ is varying slowly in comparison with the inter vortex distance, then it can be considered
constant in a uniform distribution of vortices (as a first approximation). In that case \Eqref{fullrenorm} can be
approximated by
\begin{equation}
U_\varphi\simeq\sum_{ij}\tilde\eta_{ij} u_iu_j\cos(\varphi_{ij})~\text{where}~\tilde{\eta}_{ij}=\eta_{ij}\int_{\Omega} d\mbf r  f_i(\mbf r)  f_j(\mbf r)
\label{renorm}
\end{equation}
If on the other hand $\varphi_{ij}$ varies rapidly, then it is not possible to define $\tilde{\eta}_{ij}$ without a spatial
dependence. Then $\varphi_{ij}$ will depend on $\tilde{\eta}_{ij}(\mbf r)$ which is related to the local modulation functions
$f_if_j$ and vary with a characteristic length scale.

Thus, $\tilde{\eta}$ is the effective inter-band interaction coupling resulting from density modulation.
Since in general, $f_i\not=f_j$ (unless the two bands $i,j$ are identical), one must take into account the
modulation functions $f_i$ when calculating the phase differences. In particular, if the core size in
component $i$ is larger than in component $j$, then  $\int d\mbf r f_if_k <\int d\mbf r f_jf_k$
and therefore the phase differences $\varphi_{ij}$ minimizing \Eqref{renorm}
depend on $f_i$, and consequently on the density of vortices.
Roughly speaking, introducing vortices in the system is equivalent to relative effective decrease of some of the Josephson coupling
constants.

\begin{figure}[!htb]
\hbox to \linewidth{ \hss
\includegraphics[width=0.8\linewidth]{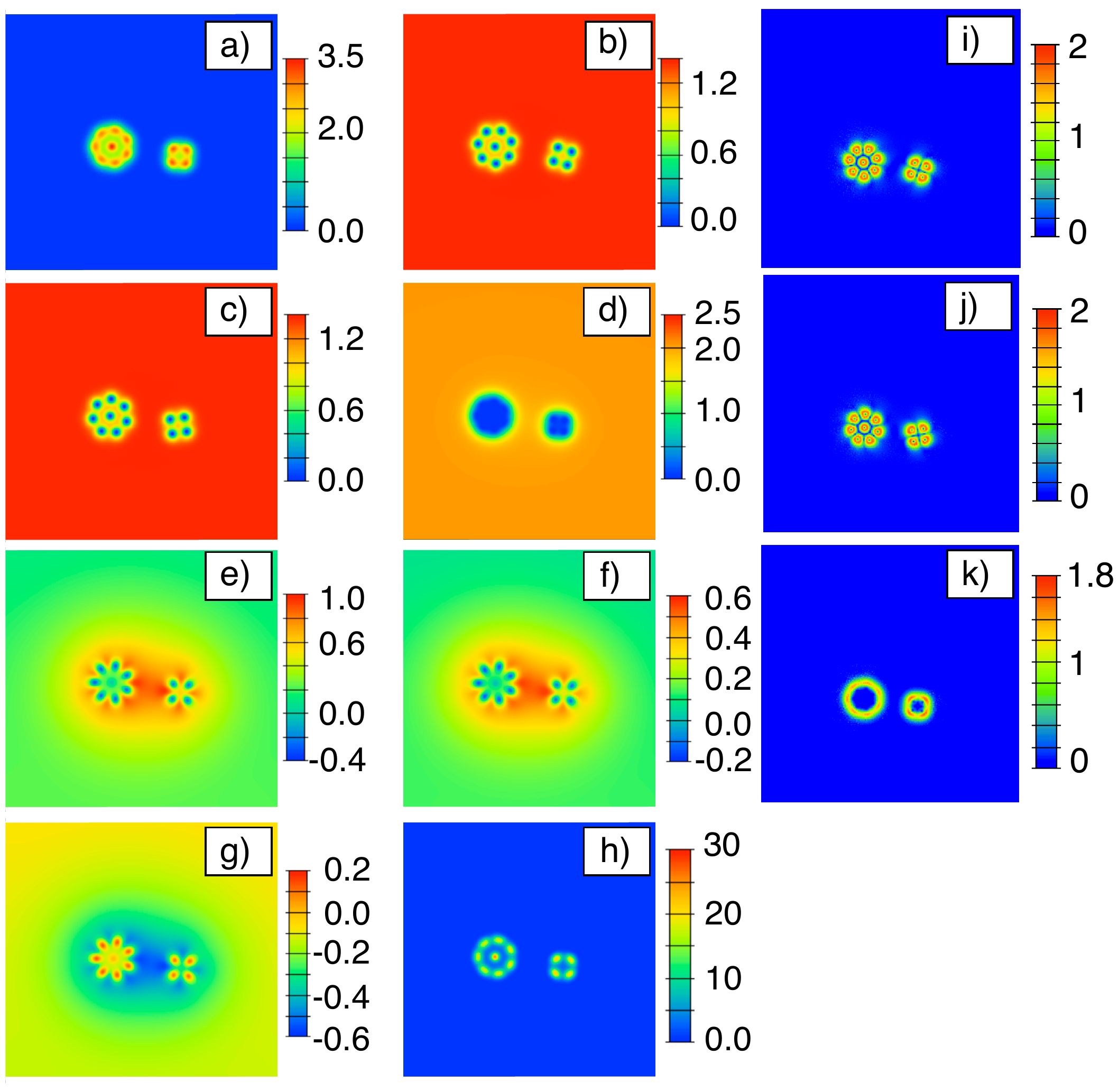}
 \hss}
\caption{
Interacting vortex clusters with internal $Z_2$ symmetry in a frustrated three band superconductor. The snapshot
represents a {non-stationary} state of the weakly interacting well-separated clusters.
In this numerical computation, each of the clusters has with a good accuracy converged to a physical solution
of GL equations, but the snapshot is taken during the slow evolution driven by  the weak
intercluster interaction. The snapshot demonstrates the existence of long-range field variations
associated with the soft mode. This produces long-range weak intervortex forces.
Displayed quantities are: a) Magnetic field, b-d) $|\psi_1|^2,|\psi_2|^2,|\psi_3|^2$, e) $|\psi_1||\psi_2|\sin\varphi_{12}$, f) $|\psi_1||\psi_3|\sin\varphi_{13})$, g) $|\psi_1||\psi_3|\sin\varphi_{23})$.
The GL parameters are $\alpha_1=-3,\;\beta_1=3,\;\alpha_2=-3,\;\beta_2=3, \alpha_3=2,\; \beta_3=0.5,\eta_{12}=2.25,\; \eta_{13}=-3.7$.
The parameter set was chosen so that it lies in the regime where the ground state symmetry of the system without
vortices is $U(1)$, but is close to the $U(1)\times Z_2$ region. Because of the disparity in vortex core size the effective interaction strengths $\tilde{\eta}_{ij}$ are depleted to different extents. As a consequence, the symmetry associated with the effective couplings inside the cluster correspond to the $U(1)\times Z_2$ regime.
}
\label{ch2}
\end{figure}

\begin{figure}[!htb]
\hbox to \linewidth{ \hss
\includegraphics[width=0.8\linewidth]{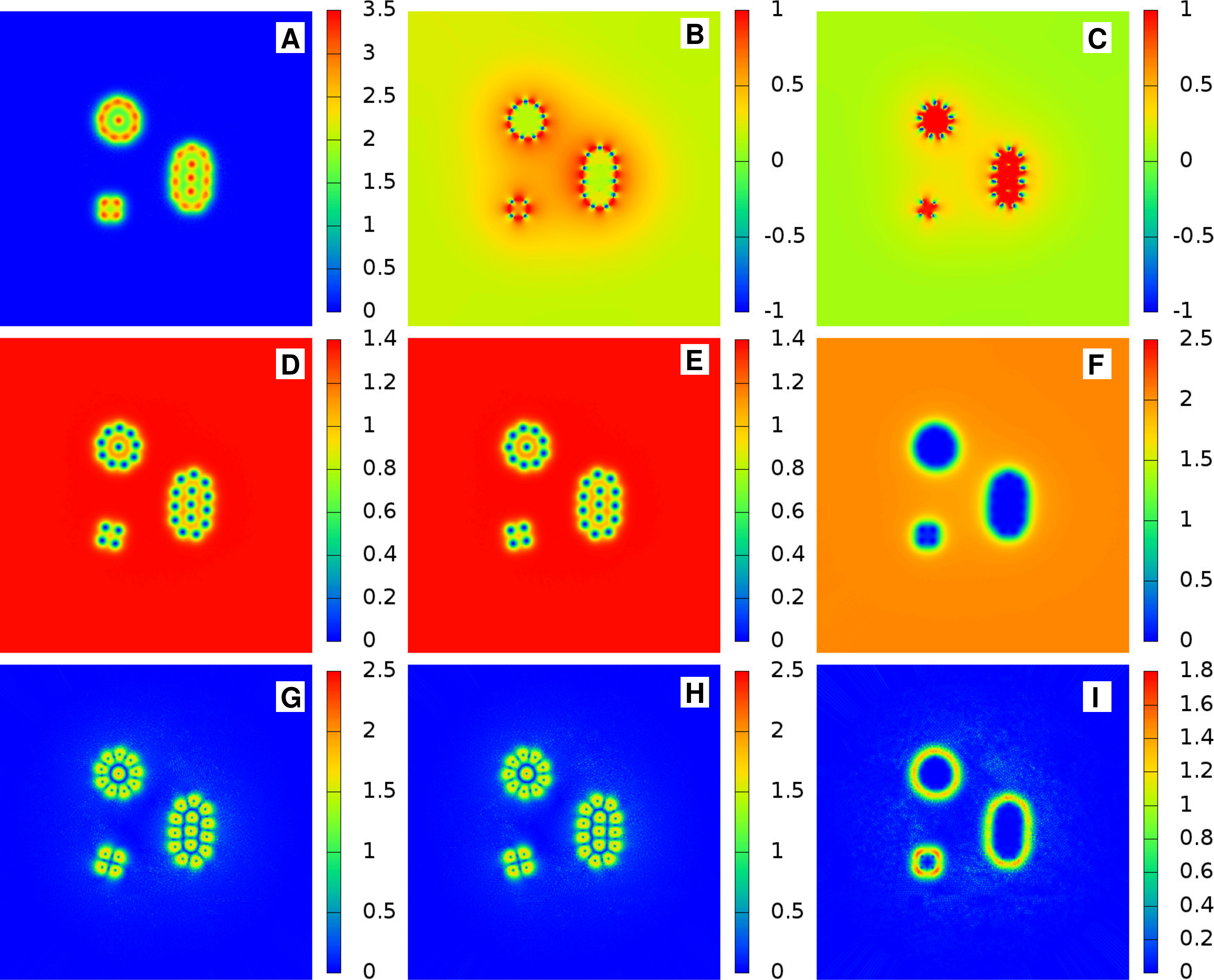}
 \hss}
\caption{
Interacting vortex clusters with internal $Z_2$ symmetry in a frustrated three band superconductor.
The panel $\mbf A$  displays the magnetic field ${B}$. Panels $\mbf B$  and $\mbf C$  respectively display $\sin\varphi_{12}$
and $\sin\varphi_{13}$, the third phase difference can obviously be obtained from these two ones. Second line,
shows the densities
of the different condensates $|\psi_1|^2$ ($\mbf D$), $|\psi_2|^2$ ($\mbf E$), $|\psi_3|^2$ ($\mbf F$). The third line displays the
supercurrent densities associated with each condensate $|J_1|$ ($\mbf G$), $|J_2|$ ($\mbf H$), $|J_3|$ ($\mbf I$). The parameter
set here is the same as in \Figref{ch2}.The instructive difference here is that th sine of the phase differences is represented
`unweighted' by the densities in contrast to \Figref{ch2}. Panel $\mbf C$ now makes clear that the inner cluster is in a defined
state $\varphi_{13} \approx \pi/2$ (whose opposite state would have been $-\pi/2$). Panel $\mbf B$ gives a visualization
of the long range interaction between the clusters.
}
\label{c32}
\end{figure}

This can have profound consequences, as the symmetry of the problem depends on the Josephson interaction terms. In Figs. \ref{ch2} , \ref{c32} we see a type-1.5 system in which the symmetry of the ground state is $U(1)$.
As vortices are inserted into the system, they form clusters and the effective inter band interactions $\tilde{\eta}_{ij}$ are renormalized to a degree
that the symmetry of the domain near vortex clusters changes to $U(1)\times Z_2$.
 Thus the Semi-Meissner state in such a system represents macroscopic phase separation in domains of broken  $U(1)$  and $U(1)\times Z_2$ symmetries.

\section{Microscopic theory of type-1.5 superconductivity}
\label{microscopic}

The phenomenological Ginzburg-Landau model described above predict the possibility
of 1.5 superconducting state. Formally the GL theory applies only at elevated temperatures.
To describe type-1.5 superconductivity in all temperature regimes (except, indeed the region
where mean-field theory is inapplicable)
as well as to
 make a connection
with a certain class of the real systems
requires a microscopic approach which also does not
rely on a GL expansion.
Such a theory was recently developed in \cite{silaev}.
So  let us consider the
described above physics in a microscopic formalism of
self-consistent Eilenberger theory. We consider a superconductor
with two overlapping bands at the Fermi level \cite{suhl}. The
corresponding two sheets of the Fermi surface are assumed to be
cylindrical. Within quasi-classical approximation the band
parameters characterizing the two different sheets of the Fermi
surface are the Fermi velocities $V_{Fj}$ and the partial
densities of states (DOS) $\nu_j$, labeled by the band index
$j=1,2$. We normalize the energies to the critical temperature
$T_c$ and length to $r_0= \hbar V_{F1}/T_c$. The system of
Eilenberger equations for two bands is
\begin{align}\label{Eq:EilenbergerF}
&v_{Fj}{\bf n_p}\left(\nabla+i {\bf A}\right) f_j +
 2\omega_n f_j - 2 \Delta_j g_j=0, \\ \nonumber
 &v_{Fj}{\bf n_p}\left(\nabla-i {\bf A}\right) f^+_j -
 2\omega_n f^+_j + 2\Delta^*_j g_j=0.
 \end{align}
 Here $\omega_n=(2n+1)\pi T$ are Matsubara frequencies and
  $v_{Fj}=V_{Fj}/V_{F1}$. The vector ${\bf n_p}=(\cos\theta_p,\sin\theta_p)$
  parameterizes the position on 2D cylindrical
 Fermi surfaces. The quasi-classical Green's functions in each band obey
 normalization condition $g_j^2+f_jf_j^+=1$. The Eilenberger differential
 Eqs.(\ref{Eq:EilenbergerF}) are solved together with the integral
 self-consistency equations for the gaps
\begin{align}\label{Eq:SelfConsistencyOP}
  \Delta_i&=T \sum_{n=0}^{N_d} \int_0^{2\pi}
 \lambda_{ij} f_j d\theta_p \\ \label{Eq:SelfConsistencyCurrent}
  {\bf j}&= -\frac{T}{\kappa^{2}}\sum_{j=1,2} n_j v_{Fj}\sum_{n=0}^{N_d}
 Im\int_0^{2\pi}  {\bf n_p} g_j d\theta_p.
\end{align}
Here $\lambda_{ij}$ is the coupling matrix which
  satisfies the symmetry relations
 $n_1\lambda_{12}=n_2\lambda_{21}$ where $n_i$ are the
 partial DOS normalized so that $n_1+n_2=1$.
We consider $\lambda_{11}>\lambda_{22}$ and
 therefore refer to the first band as ``strong" and to the second
 as ``weak".

 The asymptotics of the gap
 functions $|\Delta_{1,2}|(r)$ at distances far from the vortex core can be found linearizing the
 Eilenberger Eqs.(\ref{Eq:EilenbergerF}) together with the
 self-consistency equations.
The asymptotics of the linearized system is governed by the
singularities of response function found in Ref.(\cite{silaev})
among which are the poles and branch cuts. { In general there
are two regimes regulating the asymptotic behavior of gap
functions. The first regime is realized when two poles of the
response function lie below the branch cut.  The two poles
determine the {\it two inverse length scales} or, equivalently,
the two masses of composite gap functions fields (i.e. linear
combinations of the fields as in the previous section), which we
denote as ``heavy" $\mu_H$ and ``light" $\mu_L$ (i.e.
$\mu_H>\mu_L$). At elevated temperatures these masses are exactly
the same as the given by a corresponding two-component GL theory
obtained by the gradient expansion from the microscopic theory \cite{silaev2}.
The GL theory can be also used to describe the two massive modes
at relatively low temperatures \cite{silaev2}.
 In this case the coefficients should be
adjusted phenomenologically in order to obtain the same values of masses as the
microscopic theory yields.

 The second regime is realized at lower temperatures when there is only one pole below
the branch cut, In this case the asymptotic is determined by the
light mass mode $\mu_L$ and the contribution of the branch cut
which has all the length scales smaller than some threshold one
determined by the position of the lowest branch cut on the
imaginary axis. The branch cut contribution is essentially
non-local effect which is not captured by GL theory therefore one
can expect growing discrepancies between effective GL solution and
the result of microscopic theory at low temperatures.  }

\begin{figure}
\centerline{\includegraphics[width=0.60\linewidth]{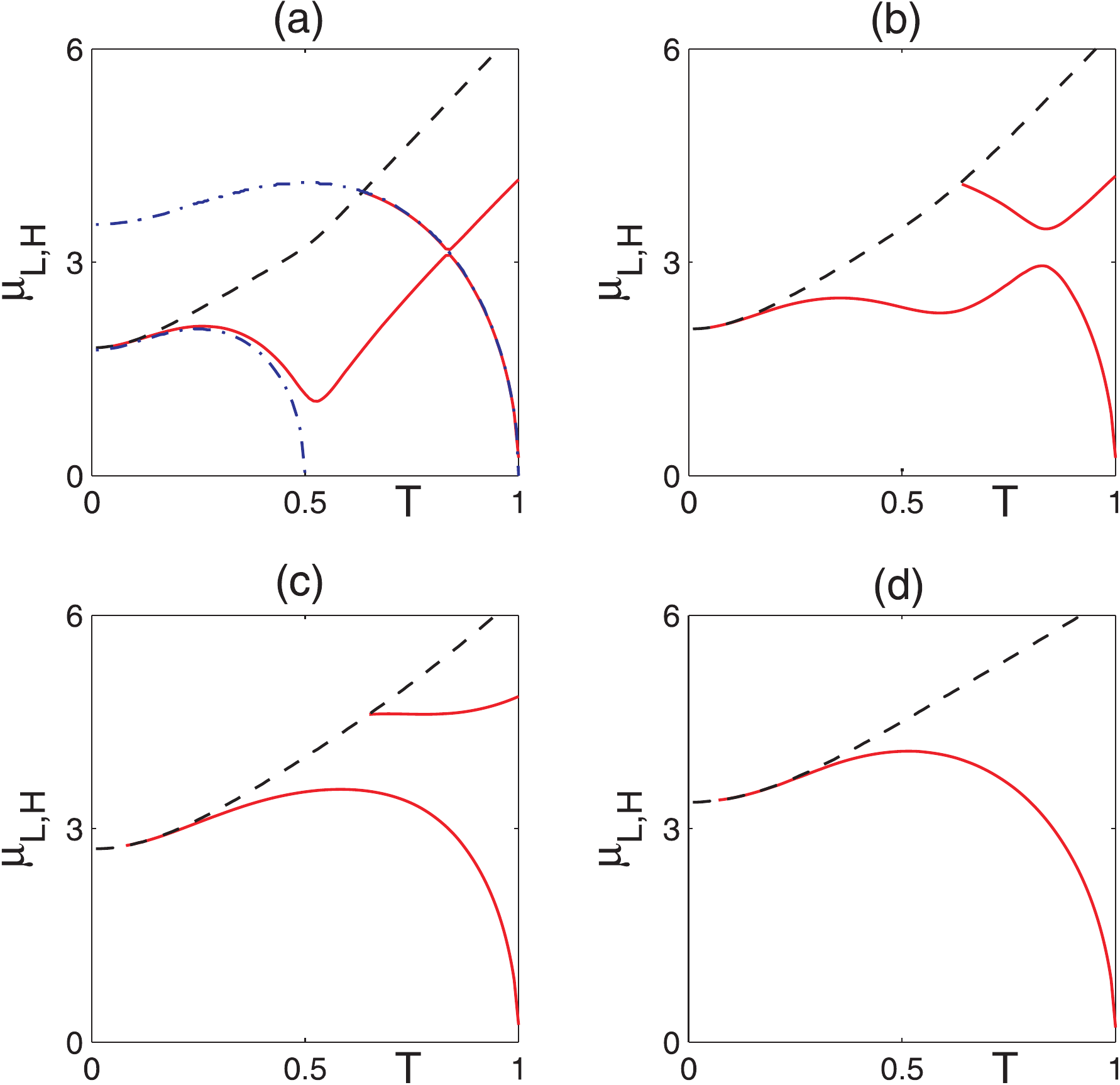}}
\caption{\label{Fig:SequenceModes} Calculated in \cite{silaev}  masses $\mu_{L}$ and $\mu_{H}$
(red solid lines) of the composite gap function fields for the
different values of interband Josephson coupling $\lambda_J$ and
$\gamma_F=1$. In the sequence of plots (a)-(d) the transformation
of masses is shown for $\lambda_J$ decreasing from the small
values $\lambda_J\ll \lambda_{11},\lambda_{22}$ to the values
comparable to intraband coupling $\lambda_J\sim
\lambda_{11},\lambda_{22}$. The particular values of coupling
constants are $\lambda_{11}=0.25$, $\lambda_{22}=0.213$ and
$\lambda_J=0.0005;\;0.0025;\;0.025;\;\lambda_{22}$ for plots (a-d)
correspondingly. By black dash-dotted lines the branch cuts are
shown. In (a) with blue dash-dotted lines the masses of modes are
shown for the case of $\lambda_J=0$. Note that at $\lambda_J=0$
the two masses go to zero at two different temperatures. Because
$1/\mu_{L,H}$ are related to the coherence length, this reflects
the fact that for $U(1)\times U(1)$ theory there are two
independently diverging coherence lengths. Note that for finite
values of interband coupling only one mass $\mu_L$ goes to zero at
one $T_c$: this is in turn a consequence of the fact that
Josephson coupling breaks the symmetry down to single $U(1)$.
  }
\end{figure}

 The examples from ref. \cite{silaev} of the temperature dependencies
 of the masses $\mu_{L,H}(T)$ are shown in the Fig.\ref{Fig:SequenceModes}.
The evolution of the masses $\mu_{L,H}$ is shown in the sequence
of plots Fig.\ref{Fig:SequenceModes}(a)-(d)  for $\lambda_J$
increasing from the small values $\lambda_J\ll
\lambda_{11},\lambda_{22}$ to the values comparable to intraband
coupling $\lambda_J\sim \lambda_{11},\lambda_{22}$.
 The two  massive modes coexist at the temperature interval
 $T^*_1<T<T_c$, where the temperature $T^*_1$
 is determined by the branch cut position, shown in the
 Fig.\ref{Fig:SequenceModes} by black dashed line. For temperatures
 $T<T^*_1$ there exists only one massive mode lying below the branch cut.
 At very low temperatures the mass $\mu_L$
 is very close to the branch cut. As the interband coupling
 parameter is increased, the temperature $T^*_1$ rises and becomes equal to $T_c$
 at some critical value of $\lambda_J=\lambda_{Jc}$.

\begin{figure}
\centerline{\includegraphics[width=0.60\linewidth]{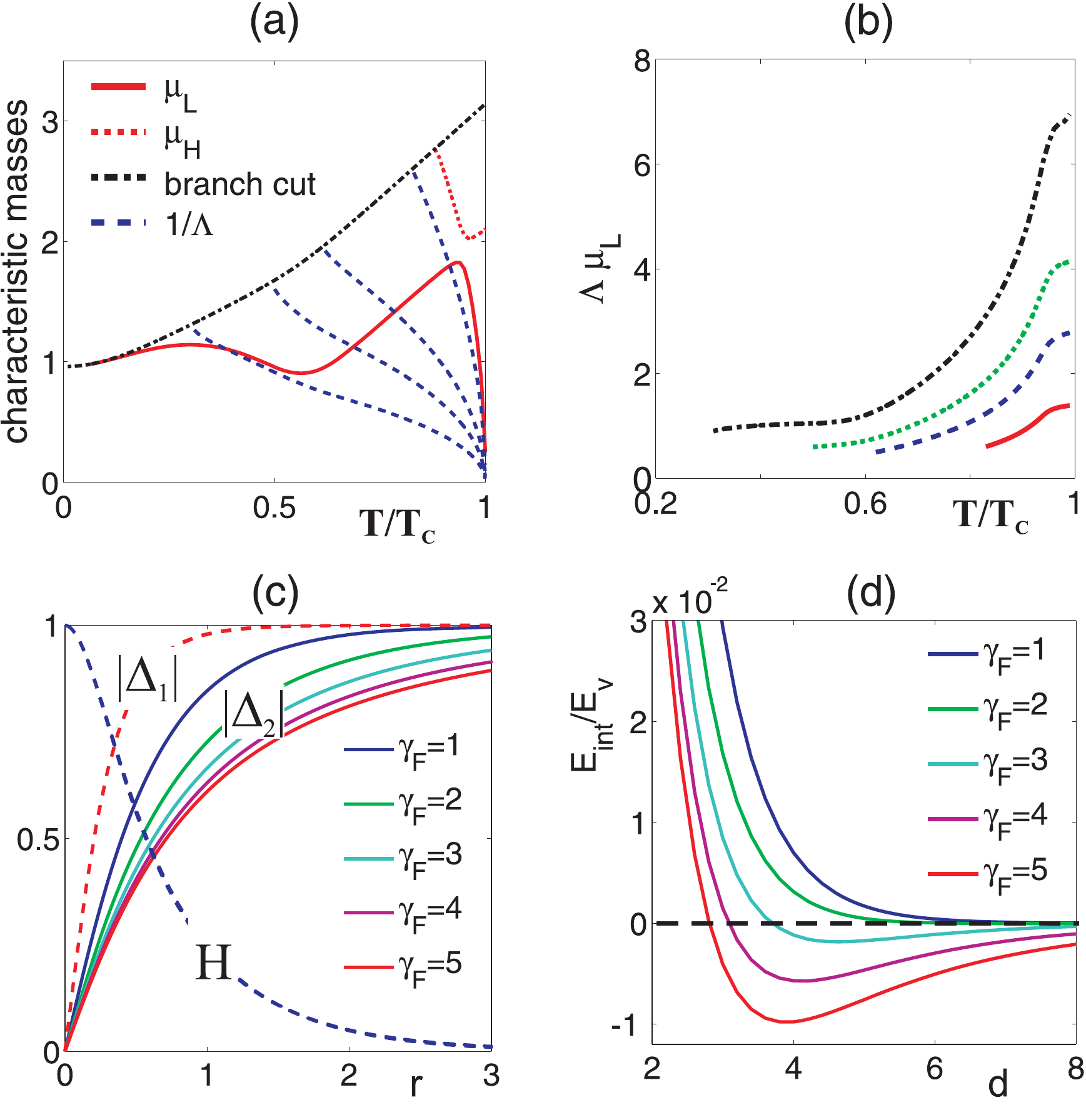}}
\caption{\label{Fig:VortexStructure15} Calculated in ref. \cite{silaev} (a) Masses $\mu_{L}$ and
$\mu_{H}$ (red solid and dotted lines) of the composite gap
function fields and inverse London penetration (blue dashed lines)
for the different values of $\Lambda \mu_L
(T_c)/\sqrt{2}=1;2;3;5$. The position of branch cut is shown by
black dash-dotted line. (b) The temperature dependence of the
quantity $\Lambda \mu_L$ for $\Lambda \mu_L(T_c)/\sqrt{2}=1;2;3;5$
(red solid, blue dashed and black dash-dotted lines). (c)
Distributions of magnetic field $H(r)/H(r=0)$, gap functions
$|\Delta_1|(r)/\Delta_{10}$ (dashed lines) and
$|\Delta_2|(r)/\Delta_{20}$ (solid lines) for the coupling
parameters $\lambda_{11}=0.25$, $\lambda_{22}=0.213$ and
$\lambda_{21}=0.0025$ and different values of the band parameter
$\gamma_F=1;2;3;4;5$. (d) The energy of interaction between two
vortices normalized to the single vortex energy as function of the
intervortex distance $d$. { In panels (c,d) the temperature is
$T=0.6$.}
 }
\end{figure}

Besides justifying the predictions of phenomenological two-component GL theory \cite{silaev2} the
microscopic formalism developed in Ref.(\cite{silaev}) allows to
describe type-1.5 superconductivity beyond the validity of GL
models. The type-1.5 behavior requires a density mode with low
mass $\mu_L$ to mediate intervortex attraction at large
separations, which should coexist with short-range repulsion. In ref. \cite{silaev}
find that the temperature dependence of $\mu_L(T)$ is
characterized by an anomalous behavior, which is in strong
contrast to temperature dependence of the mass of the  gap mode in
single-band theories. As shown on Fig.\ref{Fig:VortexStructure15}a
the function $\mu_L(T)$ is {\it non-monotonic} at low
temperatures.

Therefore for a certain range of parameters in contrast with the
physics of singe-band superconductors the product of London
penetration depth $\Lambda$ and $\mu_L$ has a strong and
nonmonotonic temperature dependence shown in
Fig.\ref{Fig:VortexStructure15}b. The inverse of  the mass of the
light composite gap mode $\mu_L$ sets the range of the attractive
density-density contribution to intervortex interaction. Therefore
the condition for the occurrence of the intervortex attraction
will be met if $\Lambda \mu_L<1$. Note that only {\it
infinitesimally} close to $T_c$, the  product can be interpreted
as single-component-like GL parameter $\kappa$ because the inverse mass
$\sqrt{2}\mu_L^{-1}$ becomes the GL coherence length.
However as shown on Fig. \ref{Fig:VortexStructure15} b) this GL parameter can have a very strong temperature
dependence, 
thus in general it {\it cannot} be used as a single parameter to characterize two-band systems.
This is because in a two-band superconductor, for a wide range
of parameters even slightly away  from $T_c$ the temperature dependence of $\mu_L$, is
dramatically different from that of the inverse magnetic field
penetration length $\Lambda^{-1}$.

 Furthermore  because the softest mode with the
mass $\mu_L$ in two band system may be associated with only a
fraction of the total condensate (as follows from corresponding
mixing angles), and because there could be the
second mixed gap mode with larger mass $\mu_H$, the short-range
intervortex interaction can be repulsive.
 In Ref.(\cite{silaev}) the temperature dependencies of $ \Lambda^{-1}$ and
$\mu_L$ are compared demonstrating how in these cases the system
 goes from type-II to type-1.5 behavior
 as temperature is decreased and
$\mu_L$ becomes smaller than  $ \Lambda^{-1}$,
 and, the density associated with the light mode is small enough that the system has
a short-range intervortex repulsion.

We calculate self-consistently the structure of isolated vortex
for different values of $\gamma_F=v_{F2}/v_{F1}$.  A complex
aspect of the vortex structure in two-band system is that in
general the exponential law of the asymptotic behavior of the gaps
is {\it not} directly related to the ``core size" at which gaps
recover most of their ground state values.  We can characterize
this effect by defining a ``healing" length $L_{\Delta i}$ of the
gap function as follows $|\Delta_i| (L_{\Delta i})= 0.95
\Delta_{i0}$. The characteristic example of the vortex structure
is shown in Fig. \ref{Fig:VortexStructure15}(c). For this case we
obtain that $L_{\Delta 1}\approx 0.8$ for all values of
$\gamma_F$. On the contrary, the healing length $L_{\Delta 2}$ of
changes significantly such that $L_{\Delta 2}= 1.6;\; 2.5;\;
3.2;\; 3.9;\;4.5$ for $\gamma_F=1;\; 2;\; 3;\; 4;\; 5$
correspondingly.

To demonstrate the type-1.5 superconductivity i.e. large-scale
attraction and small-scale repulsion of vortices which originates
from disparity of the variations of two gaps,
 the intervortex interaction energy was  calculated in \cite{silaev}.
The two-band generalization of the Eilenberger
expression was evaluated for the free energy of the two vortices positioned at
the points ${\bf r_{R}}=(d/2, 0)$ and ${\bf r_{L}}=(-d/2, 0)$ in
$xy$ plane. In Fig.\ref{Fig:VortexStructure15}(d) the interaction
energy $E_{int}$ is shown as a function of the distance between
two vortices $d$. The energy $E_{int}$ is normalized to the single
vortex energy $E_v$. The plots on
Fig.\ref{Fig:VortexStructure15}(d) clearly demonstrate the
emergence of type-1.5 behavior  when the parameter $\gamma_F$, which characterizes
the disparity in band characteristics  is
increased. This is manifested in
 the appearance non-monotonic behavior of the intervortex interaction energy $E_{int}(d)$
as a consequence of two-component structure of the theory. 

\section{Conclusion}
We reviewed the recent developments in description of  type-1.5 superconductivity in multicomponent systems.
Both at the levels of microscopic and  Ginzburg-Landau field theories
the behavior arises as a consequence of one or several of the fundamental
length scales associated with density variations $\xi_i$ being larger than the magnetic field
penetration length $\lambda$, while at the same time the system possessing thermodynamically stable
vortex excitations. These vortices have long-range attractive (originating from outer cores overlap) and short-range repulsive interaction.
This leads to an additional ``semi-Meissner" phase sandwiched between Meissner and vortex states
which is a macroscopic phase separation  into domains of Meissner and vortex states.
We discussed that in case of thermal fluctuations or/and more that two bands this phase separation
can also results in coexistence of macroscopically large domains
 with different broken symmetries: i.e.
$U(1)$ and $U(1)\times U(1)$ or $U(1)$ and $U(1)\times Z_2$.


\section{Acknowledgments}
 EB was supported by Knut and Alice Wallenberg
Foundation through the Royal Swedish Academy of Sciences, Swedish Research Council and by the US National
Science Foundation CAREER Award No. DMR-0955902.
JC was supported by the Swedish Research Council.
MS was  supported by the Swedish
Research Council, "Dynasty" foundation, Presidential RSS Council
(Grant No. MK-4211.2011.2) and Russian Foundation for Basic
Research.
JMS was supported by the UK Engineering and Physical Sciences
Research Council. EB thanks the Aspen Center for
Physics for hospitality and support under the NSF grant
No. 1066293.
The computations were performed on resources 
provided by the Swedish National Infrastructure for Computing (SNIC) 
 at National Supercomputer Center at Linkoping, Sweden.


\end{document}